\documentstyle[12pt,epsf]{article}

\begin{document}

\baselineskip=18.8pt plus 0.2pt minus 0.1pt

\makeatletter

\renewcommand{\thefootnote}{\fnsymbol{footnote}}
\newcommand{\beq}{\begin{equation}}
\newcommand{\eeq}{\end{equation}}
\newcommand{\bea}{\begin{eqnarray}}
\newcommand{\eea}{\end{eqnarray}}
\newcommand{\nn}{\nonumber}
\newcommand{\hs}[1]{\hspace{#1}}
\newcommand{\vs}[1]{\vspace{#1}}
\newcommand{\Half}{\frac{1}{2}}
\newcommand{\p}{\partial}
\newcommand{\ol}{\overline}
\newcommand{\wt}[1]{\widetilde{#1}}
\newcommand{\ap}{\alpha'}
\newcommand{\bra}[1]{\left\langle  #1 \right\vert }
\newcommand{\ket}[1]{\left\vert #1 \right\rangle }
\newcommand{\vev}[1]{\left\langle  #1 \right\rangle }

\newcommand{\ul}[1]{\underline{#1}}

\makeatother

\begin{titlepage}
\title{
\hfill\parbox{4cm}
{\normalsize KUNS-1680\\{\tt hep-th/0008247}}\\
\vspace{1cm}
The M2-brane Soliton on the M5-brane with Constant 3-Form
}
\author{Yoji Michishita
\thanks{
{\tt michishi@gauge.scphys.kyoto-u.ac.jp}
}
\\[7pt]
{\it Department of Physics, Kyoto University, Kyoto 606-8502, Japan}
}

\date{\normalsize August, 2000}
\maketitle
\thispagestyle{empty}

\begin{abstract}
\normalsize
 We obtain a BPS soliton of the effective theory of the M5-brane
worldvolume with constant 3-form representing M2-branes ending on the
M5-brane. The dimensional reduction of this solution agrees with the
 known results on D-branes.
 \end{abstract}
\end{titlepage}

Recently physics on the D-brane in constant NS-NS 2-form 
($B_{\mu\nu}$)
background is extensively studied. The worldvolume theory on the 
D-brane is described by either ordinary gauge theory or 
noncommutative gauge theory \cite{sw}. Furthermore the
lift to M-theory is considered \cite{gmss,bbss1,bbss2}. It is the
theory on the M5-brane with constant 3-form. In this theory M2-branes
ending on the M5-brane play important roles.
The worldvolume theory on the D-brane with constant $B_{\mu\nu}$ has
 various solitons, which are interpreted as fundamental strings or 
D-branes ending on it. They are tilted with some angle determined by
$B_{\mu\nu}$. Therefore it is natural to consider such a kind of 
solitons on the M5-brane i.e. M2-branes ending on the M5-brane with
constant 3-form.
The purpose of this paper is to obtain the BPS solution of the
equations of motion representing such a configuration.
 The case without 3-form is considered in \cite{hlw}, 
and on special Lagrangian submanifolds with constant 3-form in
\cite{lm}. 

There are several different formulations for the effective theory of 
M5-brane: the covariant field equations from the superembedding
approach \cite{hsw}, the noncovariant Lagrangian \cite{apps}, the
covariant Lagrangian with an auxiliary field \cite{blnpst1,blnpst2},
the formulation constructed in \cite{cns}, etc.
The first one leads to the same field equations as those from the
third one \cite{blnpst3} and the second one is
equivalent to the third 
one with the appropriate gauge fixing \cite{apps}. The fourth one is
also shown to be equivalent to the others \cite{cns}.
Therefore we 
can use any of these formulations for seeking solutions of the
field equations. In this paper we use the first formulation.

First we will explain the notation. On the M5-brane worldvolume
tangent space indices are denoted by $a,b,\dots=0,1,2,\dots,5$ and 
world indices by $m,n,\dots=0,1,2,\dots,5$. Indices of the target 
space are denoted by the same symbols with underline.
The bosonic fields on the M5-brane are $X^{\ul{m}}(\sigma)$ and
$A_{mn}$, where $X^{\ul{m}}(\sigma)$ is the embedding map from the
worldvolume into 
the 1+10 dimensional spacetime and $A_{mn}$ is the 2-form which 
satisfies the nonlinear selfduality condition explained below. 
The field strength of $A_{mn}$ is always appeared in the combination
$H_{mnp}=3\p_{[ m}A_{np] }+C_{mnp}$ because of the gauge invariance
for the bulk 3-form $C_{\ul{m}\ul{n}\ul{p}}$. ($C_{mnp}$ is the
pullback of $C_{\ul{m}\ul{n}\ul{p}}$.)

In this paper, we mainly consider the case where the bulk metric is 
flat and $C_{\ul{m}\ul{n}\ul{p}}$ is constant, and fermions on the 
M5 brane are taken 
to be zero since we are interested in classical solutions.
The induced worldvolume metric $g_{mn}$ and vielbein $e_m^a$ are 
defined by
\beq
g_{mn}=\eta_{ab}e_m^ae_n^b=\p_m X^{\ul{m}}\p_n X^{\ul{n}}
 \eta_{\ul{a}\ul{b}}E_{\ul{m}}^{\ul{a}}E_{\ul{n}}^{\ul{b}} .
\eeq
We introduce the auxiliary 3-form $h_{abc}$ satisfying the following 
selfdual condition.
\beq
h^{abc}=\frac{1}{6}\epsilon^{abcdef}h_{def} ,
\eeq
$H_{mnp}$ is related to $h_{abc}$ by $H_{mnp}=e_m^a(m^{-1})_b^a
e_n^ce_p^d h_{bcd}$ and $m_a^b\equiv\delta_a^b-2h_{acd}h^{bcd}$. 
The $\kappa$-symmetry of the M5-brane worldvolume theory is $\delta 
\theta = \kappa(1+\Gamma)$ and
\beq
\Gamma=\frac{1}{6!\sqrt{-g}}\epsilon_{abcdef}\gamma^{abcdef}
+\frac{1}{3}h_{abc}\gamma^{abc} ,
\eeq
where $\gamma_a\equiv e_a^m\p_m
X^{\ul{m}}E_{\ul{m}}^{\ul{a}}\Gamma_{\ul{a}}$
and $\Gamma_{\ul{a}}$ are the 10+1 dimensional gamma matrices.
Hereafter we take the static gauge $X^0,\cdots,X^5=\sigma^0,
\cdots,\sigma^5$ and other $X^{\ul{m}}$ are taken zero except 
$X\equiv X^6$.
The equations of motion for $X$ and $A_{mn}$ are given by
\bea
G^{mn}\nabla_m\nabla_nX & = & 0 , \\
G^{mn}\nabla_mH_{npq} & = & 0 ,
\eea
where $G^{mn}=e_a^m(m^2)^a_be^{bn}$, and 
$\nabla_m$ is the covariant derivative with the Christoffel
symbol $\Gamma_{mn}^p=\p_m(\p_nX^{\ul{r}}E_{\ul{r}}^{\ul{c}})
\p_qX^{\ul{s}}E_{\ul{s}\ul{c}}g^{qp}$.
If we take $H_{mnp}$ constant, then in the generic case we can take all 
the components except $h_{012}$ and $h_{345}$ zero by choosing 
appropriate coordinate \cite{sw,bbss1}. 
\footnote{There is a nongeneric case where $h_{012}= -h_{512}=
-h_{034}= h_{534}\equiv h$ \cite{sw,bbss1}. 
We can obtain this from the generic 
case by infinite boost in the 0-5 direction and taking the
limit $h\rightarrow 0$.}

\begin{figure}[htdp]
\begin{center}
\leavevmode
\epsfxsize=70mm
\epsfbox{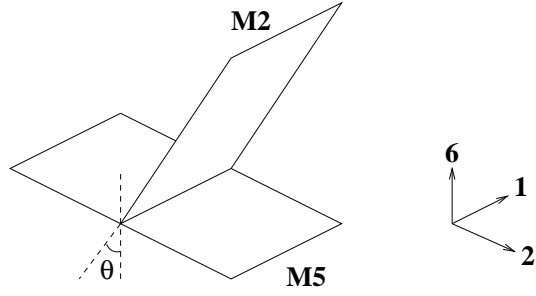}
\caption{the M2-brane ending on the M5-brane}  
\label{fig1}
\end{center}
\end{figure}

Let us consider the case where the M5-brane lies in the 
directions $X^{0,1,2,3,4,5}$ i.e. $X=0$ and 
$h_{012}=-h_{345}\equiv h=$ constant with other components
vanishing. Then,
\beq
\Gamma=-\Gamma^{012345}+2h(\Gamma^{012}-\Gamma^{345}) .
\eeq
The unbroken supersymmetry parameter $\xi$ satisfies
\beq
\xi= \Gamma \xi .
\label{m5susy}
\eeq
In addition, let us consider a M2-brane ending on the M5-brane, and 
investigate the BPS configuration.
From the case of the solitons on D-branes, we can expect that the 
M2-brane is tilted. Therefore we assume that it lies in the 
directions $X^0,X^1$, and $\cos\theta X^6 + \sin\theta X^2$.
See Fig. \ref{fig1}.
Hence
\beq
\xi= \eta_1 \Gamma^0\Gamma^1(\cos\theta\Gamma^6+
 \sin\theta\Gamma^2) \xi\equiv\eta_1\Gamma'\xi .
\label{m2susy}
\eeq
(Note that the $\kappa$-symmetry of the M2-brane, and therefore 
the unbroken supersymmetry, are independent 
of the 3-form.) 
Here $\eta_1=\pm 1$ corresponds to the charge of the M2-brane, i.e.
M2 or anti-M2. It is well 
known that $\theta=0$ without the 3-form is a BPS configuration.
Then
\beq
\Gamma \xi=  [2\eta_1h\sin\theta+2\eta_1h\cos\theta\Gamma^{26}
 +(-2h+\eta_1\sin\theta)\Gamma^{345}
 -\eta_1\cos\theta\Gamma^{23456}]\xi .
\eeq
If we impose $\xi$ the condition
\beq
\xi=\eta_2\Gamma^{23456}\xi ,
\label{adsusy}
\eeq
and $\eta_2=\pm 1$, then
\beq
\Gamma \xi= [(2\eta_1h\sin\theta-\eta_1\eta_2\cos\theta)
 +(2\eta_1\eta_2h\cos\theta-2h+\eta_1\sin\theta)
 \Gamma^{345}]\xi .
\eeq
Hence there are some unbroken supersymmetries if we require
\beq
h=-\Half\frac{\eta_2\sin\theta}{\cos\theta-\eta_1\eta_2} .
\label{voh}
\eeq
Thus the independent conditions for $\xi$ are eq.(\ref{m2susy}) 
and (\ref{adsusy}) and eq.(\ref{m5susy}) is derived from them.
This configuration preserves $\frac{1}{4}$ supersymmetry.
If we require that $h\rightarrow 0$ corresponds to 
$\theta\rightarrow 0$ we obtain $\eta_1\eta_2=-1$.
The nonzero component of $H_{mnp}$ is
\bea
H_{012} & =  & \frac{1}{4}\eta_1\sin\theta , \label{voh2}\\
H_{345} & =  & \frac{1}{4}\eta_2\tan\theta . \label{voh3}
\eea

Next we consider a probe M2-brane in the background representing the
M5-brane with constant 3-form found in \cite{rtcghn}:
\bea
ds_{11} & = & k^{1/3}f^{1/3}[ f^{-1}(-(dX^0)^2+(dX^1)^2+(dX^2)^2)
 +k^{-1}((dX^3)^2+(dX^4)^2+(dX^5)^2)\nn\\
& & +(dX^6)^2+\cdots +(dX^{10})^2] , \nn\\
dC & = & \frac{1}{4}(\sin\chi df^{-1}
 \wedge dX^0\wedge dX^1\wedge dX^2
 -\tan\chi dk^{-1}\wedge dX^3\wedge dX^4\wedge dX^5 \nn\\
& & +\frac{1}{4!}\cos\chi \epsilon_{ijklm}\p_mfdx^idx^jdx^kdx^l) , \nn\\
& & f=1+\frac{R^3}{r^3},\quad k=\sin^2\chi+\cos^2\chi f , \nn\\
& & i,j,\cdots=6,7,\cdots,10 ,
\eea
and investigate the BPS configuration of the M2-brane to compare with
the previous result.
We take the following ansatz for the embedding map from the
M2-brane worldvolume into the 1+10 dimensional spacetime.
\bea
X^0 & = & \sigma^0 , \nn\\
X^1 & = & \sigma^1 , \nn\\
\sin\theta X^2+\cos\theta X^6 & = & \sigma^2 , \nn\\
\cos\theta X^2-\sin\theta X^6 & = & 0 , \nn\\
X^{3,4,\cdots,10} & = & 0 .
\label{com2}
\eea
Then the bosonic part of the effective action of the M2-brane is
\bea
S & = & -T_{M2}\int d^3\sigma[\sqrt{-det(\p_mX^{\ul{m}}\p_nX^{\ul{n}}
 E_{\ul{m}}^{\ul{a}}E_{\ul{n}\ul{a}})}-\eta_1
 \p_0X^{\ul{m}}\p_1X^{\ul{n}}\p_2X^{\ul{p}}C_{\ul{m}\ul{n}\ul{p}}]\nn\\
& = & -T_{M2}\int d^3\sigma[k^{1/2}f^{-1}\sqrt{\sin^2\theta+
 \cos^2\theta f}-\eta_1\sin\chi\sin\theta f^{-1}] .
\eea
Here we use $C_{012}=\frac{1}{4}\sin\chi f^{-1}$ 
and $C_{345}=-\frac{1}{4}\tan\chi k^{-1}$.
If we take $\chi=\eta_1\theta$, then
\beq
S = -T_{M2}\int d^3\sigma \cos^2\chi ,
\eeq
and the M2-brane has no potential term for fluctuations. This shows
that this configuration is BPS. We identify the value of $C_{012}$ and
$C_{345}$ at the infinity of the space transverse to the M5-brane with
3-form flux on the M5-brane, as was done in 
\cite{bs}:
\beq
H_{012}=\frac{1}{4}\sin\chi=\frac{1}{4}\eta_1\sin\theta, \quad 
H_{345}=-\frac{1}{4}\tan\chi=-\frac{1}{4}\eta_1\tan\theta.
\eeq
This value and the configuration of M2-brane (\ref{com2}) near the
infinity of the space transverse to the M5-brane
are consistent with the previous result in the flat background.

Now we solve the equations of motion of the M5-brane to obtain the BPS
solution 
representing the above configuration in the flat background.
We consider the same ansatz as in \cite{hlw}:
\bea
h_{01a'} & = & v_a , \nn\\
h_{a'b'c'} & = & \epsilon_{a'b'c'd'}v^{d'} .
\eea
with the other components of $h_{abc}$ vanishing, and $X$ and $v_{a'}$
are independent of $\sigma^0$ and $\sigma^1$. 
Here primed indices run $2,\cdots,5$. 
Then
\bea
m_a^b & = & \left(\begin{array}{ccc}
 1+4v^2 & & \\
 & 1+4v^2 & \\
 & & (1-4v^2)\delta_{a'}^{b'}+8v_{a'}v^{b'}
 \end{array}\right) , \\
g_{mn} & = & \left(\begin{array}{ccc}
 -1 & & \\
 & 1 & \\
 & & \delta_{m'n'}+\p_{m'}X\p_{n'}X
 \end{array}\right) , \\
e_m^a & = &  \left(\begin{array}{ccc}
 1 & & \\
 & 1 & \\
 & & \delta_{m'n'}+c\p_{m'}X\p_{n'}X
 \end{array}\right) , \\
e_a^m & = &  \left(\begin{array}{ccc}
 1 & & \\
 & 1 & \\
 & & \delta_{m'n'}+c'\p_{m'}X\p_{n'}X
 \end{array}\right) , \\
G^{mn} & = & \left(\begin{array}{ccc}
 -(1+4v^2)^2 & & \\
 & (1+4v^2)^2 & \\
 & & (1-4v^2)^2g^{m'n'}+16v^{a'}e_{a'}^{m'}v^{b'}e_{b'}^{n'}
 \end{array}\right) .
\eea
From $g_{mn}=e_m^ae_{na}$ and $g^{mn}=e_a^me^{an}$ we can determine
$c$ and $c'$: $c=(\p X)^{-2}(-1\pm\sqrt{1+(\p X)})$,
$c'=(\p X)^{-2}(-1\pm\frac{1}{\sqrt{1+(\p X)^2}})$ and 
$(\p X)^2\equiv\p_{m'}X\p_{n'}X\delta^{m'n'}$. We take the branch 
which is smoothly connected to the case $X=0$: 
$c=(\p X)^{-2}(-1+\sqrt{1+(\p X)})$,
$c'=(\p X)^{-2}(-1+\frac{1}{\sqrt{1+(\p X)^2}})$.
The nonvanishing components of $H_{mnp}$ are
\bea
H_{01m'} & = & \frac{1}{1+4v^2}e_{m'}^{a'}v_{a'} , \\
H_{m'n'p'} & = & \frac{\sqrt{1+(\p X)^2}}{1-4v^2}\epsilon_{m'n'p'q'}
 e_{a'}^{q'}v^{a'} .
\eea
Let us consider the BPS condition. The unbroken supersymmetry 
parameter $\xi$ satisfies $\Gamma\xi=\xi$ and
\bea
\Gamma & = & \frac{1}{\sqrt{1+(\p X)^2}}(-\Gamma^{012345}+\p_{a'} X 
 \Gamma^{a'} \Gamma^{016}\Gamma^{2345}) 
 + 2(v_{a'}+c'(v\p X)\p_{a'} X)\Gamma^{01a'} \nn\\
& & -2\epsilon v_{a'}\Gamma^a\Gamma^{2345}
 -2c'((\p X)^2v_{a'}-(v\p X)\p_{a'} X)\Gamma^{a'}\Gamma^{2345} 
 \nn\\
& & + 2\frac{1}{\sqrt{1+(\p X)^2}}(v\p X)\Gamma^{019}
 + \frac{1}{\sqrt{1+(\p X)^2}}\epsilon_{a'b'c'd'}
 \p^{a'}v^{d'}\Gamma^{b'c'9} ,
\eea
where $(v\p X)=v_{a'}\p_{m'}X\delta^{a'm'}$.
$\xi$ should satisfy eq.(\ref{m5susy}) and (\ref{m2susy}).
We impose the condition (\ref{m2susy}) and (\ref{adsusy}) 
as in the discussion above and require that eq.(\ref{m5susy})
is satisfied.
\bea
\Gamma\xi & = & \left[ \frac{1}{\sqrt{1+(\p X)^2}}\eta_1(-\eta_2
 +2(v\p X))\cos\theta
 +\eta_1\sin\theta(\eta_2\frac{1}{\sqrt{1+(\p X)^2}}\p_2X
  +2c'(v\p X)\p_2 X +2v_2) \right]\xi \nn\\
& & +\Biggl[-\frac{1}{\sqrt{1+(\p X)^2}}\eta_1(-\eta_2+2(v\p X))
 \sin\theta 
 + \eta_1\cos\theta(\eta_2\frac{1}{\sqrt{1+(\p X)^2}}
  \p_2 X +2c'(v\p X)\p_2X +2v_2) \nn\\
& & -2\eta_2(\frac{1}{\sqrt{1+(\p X)^2}}v_2
  -c'(v\p X)\p_2X)\Biggl]\Gamma^{26}\xi \nn\\
& & +\left[\eta_1\cos\theta(\eta_2\frac{1}{\sqrt{1+(\p X)^2}}
 \p_\alpha X +2c'(v\p X)v_\alpha+2v_\alpha)
 -2\eta_2(\frac{1}{\sqrt{1+(\p X)^2}}v_\alpha
 -c'(v\p X)\p_\alpha X)
 \right]\Gamma^{\alpha 6}\xi \nn\\
& & +\left[\eta_1\sin\theta(\eta_2\frac{1}{\sqrt{1+(\p X)^2}}
 \p_\alpha X +2c'(v\p X)v_\alpha+2v_\alpha)
 +2\eta_2\frac{1}{\sqrt{1+(\p X)^2}}
  (\p_2Xv_\alpha-v_2\p_\alpha X)
 \right]\Gamma^{\alpha 2}\xi \nn\\
& & +2\eta_2\frac{1}{\sqrt{1+(\p X)^2}}v_\alpha \p_\beta
 X\Gamma^{\alpha\beta}\xi .
\eea
Here $\alpha,\beta,\cdots$ run $3,4,5$.
To satisfy eq.(\ref{m5susy}), we take
the coefficient of $\Gamma^{\alpha 6}$ and $\Gamma^{\alpha 2}$
zero and obtain the following BPS equations.
\bea
v_2 & = & \frac{\eta_1\eta_2}{2}\frac{\sin\theta+\cos\theta\p_2X
 +\sin\theta\frac{\p_\alpha X\p^\alpha X}{(\p X)^2}}
 {-\eta_1\cos\theta+\eta_1\sin\theta\p_2 X +\eta_2\sqrt{1+(\p X)^2}} ,
 \label{bps1}\\
v_\alpha & = & \frac{\eta_1\eta_2}{2}\frac{\cos\theta
 +\frac{1}{(\p X)^2}\left(1-\frac{1}{\sqrt{1+(\p X)^2}}\right)
 \sin\theta\p_2X}
 {-\eta_1\cos\theta+\eta_1\sin\theta\p_2 X +\eta_2\sqrt{1+(\p X)^2}}
 \p_\alpha X .
 \label{bps2}
\eea
then by straightforward calculation
\beq
\Gamma\xi=\xi ,
\eeq
i.e. eq.(\ref{m5susy}) is satisfied.

The nonzero components of $H_{mnp}$ under the BPS condition are
\bea
H_{012} & = & \frac{\eta_1}{4}(\sin\theta+\cos\theta\p_2 X)
 , \\
H_{01\alpha} & = & \frac{\eta_1}{4}\cos\theta\p_\alpha X
 , \\
H_{\alpha\beta\gamma} & = & \frac{\eta_2}{4}
 \epsilon_{\alpha\beta\gamma}\left(\p_2 X
 +\frac{\sin\theta(1+(\p X)^2)}{\cos\theta-\sin\theta\p_2 X}\right)
 , \\
H_{2\alpha\beta} & = & -\frac{\eta_2}{4}
 \epsilon_{\alpha\beta\gamma}\p^\gamma X .
\eea
And 
\bea
G^{\alpha\beta} 
 & = & 4\delta^{\alpha\beta}\frac{(\cos\theta-\sin\theta\p_2 X)^2}
 {[ -\eta_1\cos\theta+\sin\theta\p_2X
 +\eta_2\sqrt{1+(\p X)^2} ]^2} , \\
G^{2\alpha} 
 & = & 4\p_{\alpha}X\frac{\sin\theta(\cos\theta-\sin\theta\p_2 X)}
 {[ -\eta_1\cos\theta+\eta_1\sin\theta\p_2X
 +\eta_2\sqrt{1+(\p X)^2} ]^2} , \\
G^{22} & = & 4\frac{1-\sin^2\theta\p_\alpha X\p^\alpha X}
 {[ -\eta_1\cos\theta+\eta_1\sin\theta\p_2X
 +\eta_2\sqrt{1+(\p X)^2} ]^2} .
\eea
The equation of motion for $X$ is
\beq
G^{m'n'}\nabla_{m'}\p_{n'}X = \frac{1}{1+(\p X)^2}G^{m'n'}
 \p_{m'}\p_{n'}X=0 .
\label{eqofx1}
\eeq
This equation is rewritten as follows.
\beq
\p_\alpha\p^\alpha X+\p_2\p^2X +\p_2\left(\frac{\sin\theta(1+(\p X)^2)}
{\cos\theta-\sin\theta\p_2 X}\right) =0 .
\label{eqofx2}
\eeq
Using eq.(\ref{eqofx1}) or (\ref{eqofx2}), the nontrivial
components of the equation of motion for $H_{mnp}$ are
\bea
G^{m'n'}\nabla_{m'}H_{n'01} & = & G^{m'n'}\p_{m'}H_{n'01} = 0 ,\\
G^{m'n'}\nabla_{m'}H_{n'p'q'} & = & G^{m'n'}(\p_{m'}H_{n'p'q'}
 -\p_{m'}\p_{p'}X\p_{r'}X g^{r's'}H_{n's'q'}
 -\p_{m'}\p_{q'}X\p_{r'}Xg^{r's'}H_{n'p's'}) \nn\\
& = & 0 .
\eea
It can be shown that these are satisfied
by using eq.(\ref{eqofx1}) or (\ref{eqofx2}) further. Furthermore the
Bianchi identity
\beq
\p_{[m}H_{npq]}=0
\eeq
is also satisfied. Hence all we should do is to solve
eq.(\ref{eqofx1}) or (\ref{eqofx2}). 

Solving eq.(\ref{eqofx2}) seems very difficult. However, since we
expect that the 
M2-brane is tilted, we can simplyfy this equation by using the rotated
coordinate $(\bar{X},\bar{z})$, as was done in \cite{m}. 
$(\bar{X},\bar{z})$ is related to $(X,z\equiv\sigma^2)$ as follows.
\beq
\left(\begin{array}{c}
 \bar{X}\\ \bar{z} \end{array}\right)
= \left(\begin{array}{cc}
 \cos\theta & \sin\theta \\
 -\sin\theta & \cos\theta
 \end{array}\right)
 \left(\begin{array}{c}
 X \\ z
 \end{array}\right) .
\eeq
We denote the derivatives with respect to $\bar{z}$ and 
$\sigma_\alpha$ by $\bar{\p}_z$ and $\bar{\p}_\alpha$.
Then
\bea
\p_z\bar{z} & = & \frac{1}{\cos\theta+\sin\theta\bar{\p}_z\bar{X}} , \\
\p_\alpha\bar{z} & = & -\frac{\sin\theta\bar{\p}_\alpha\bar{X}}
 {\cos\theta+\sin\theta\bar{\p}_z\bar{X}} .
\eea
Eq.(\ref{eqofx2}) is rewritten as follows.
\beq
\bar{\p}_\alpha\bar{\p}^\alpha\bar{X}+\bar{\p_z}\bar{\p_z}\bar{X} =0 .
\eeq
This can be easily solved. The solution with the boundary condition
that the M5-brane worldvolume lies in the direction $X^{0,1,2,3,4,5}$
at the infinity is 
\beq
\bar{X} = \sum_i\frac{Q_i}{(\bar{z}^2+(\sigma^\alpha)^2)
 -(\bar{z}_i^2+(\sigma^\alpha_i)^2)}
 +\tan\theta\bar{z} .
\label{sol}
\eeq
\begin{figure}[htdp]
\begin{center}
\leavevmode
\epsfxsize=70mm
\epsfbox{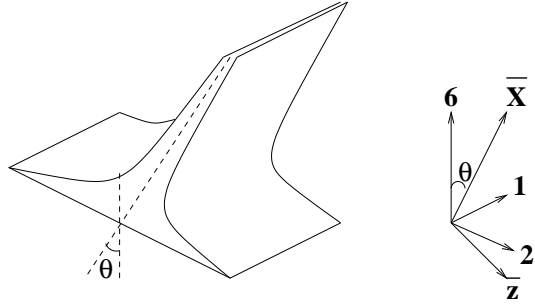}
\caption{the solution on the M5-brane}  
\label{fig2}
\end{center}
\end{figure}
We can interpret this solution as tilted M2-branes emanating from
$(\bar{z}_i, \sigma^\alpha_i)$. See Fig \ref{fig2}.
From this solution we can obtain the values of $h_{abc}$ at the
infinity of the worldvolume:
\bea
h_{012} & \rightarrow & -\Half\frac{\eta_2\sin\theta}
{\cos\theta-\eta_1\eta_2} , \\
h_{345} & \rightarrow & \Half\frac{\eta_2\sin\theta}
{\cos\theta-\eta_1\eta_2} , \\
H_{012} & \rightarrow & \frac{1}{4}\eta_1\sin\theta , \\
H_{345} & \rightarrow & \frac{1}{4}\eta_2\tan\theta ,
\eea
with the other components vanishing. These values agrees with 
eq.(\ref{voh}), (\ref{voh2}), and (\ref{voh3}).
The coefficients $Q_i$ are given in (\ref{charge}) below.

There is the critical value for $H_{012}$ i.e. 
$H_{012}=\pm\frac{1}{4}$ \cite{gmss,bbss2}. At this value 
$\theta=\pm\frac{\pi}{2}$ and the M2-brane is parallel to the
M5-brane.

Next we consider the dimensional reduction of this solution.
If we compactify the direction $m$ along the M5-brane, we have a D4
brane in type IIA string theory and \cite{hlw}
\beq
H_{mnp}=\frac{1}{4}(F_{np}+B_{np}) ,
\eeq
where $F_{np}$ is the field strength of the gauge field on the
D4-brane and $B_{np}$ is (pullback of) the NS-NS 2-form.
If we compactify the direction perpendicular to the M5-brane,
we obtain a NS5-brane in a constant RR 3-form.
Let us consider three cases.
The first case is the double dimensional reduction along
the direction $X^1$. We obtain the configuration that tilted F-strings
ending on a D4-brane in constant $B_{02}$. The relationship between
the constant NS-NS 2-form and the angle $\theta$ is
\beq
B_{02} =\eta_1\sin\theta ,
\eeq
and the solution is in the same form as eq.(\ref{sol}). This result
agrees with that of \cite{h}.
We can determine the coefficient $Q_i$ from the argument in \cite{cm}:
\beq 
Q_i=\pm\frac{4\pi^3\ap^{3/2}g_s}{\Omega_3}
 =\pm\frac{4\pi^3\ell_p^3}{\Omega_3} ,
\label{charge}
\eeq
where $g_s$ is the string coupling constant, $\ell_p$ is the 11
dimensional Planck length and $\Omega_3$ is the volume of $S^3$.
The second case is the double dimensional reduction along
the direction $X^5$. We obtain the configuration that tilted D2-branes 
ending on a D4-brane in constant $B_{34}$:
\beq
B_{34} =\eta_2\tan\theta ,
\eeq
and the solution is in the same form as eq.(\ref{sol}). This result
agrees with that of \cite{m}.

The third case is the dimensional reduction along
the direction $X^{10}$. We obtain the configuration that tilted
D2-branes ending on a NS5-brane in constant RR 3-form.
The geometry of the solution is the same as in the 11 dimensional
case.

\vs{.5cm}
\noindent
{\large\bf Acknowledgments}\\[.2cm]
I would like to thank S.\ Moriyama and P.\ K.\ Townsend for helpful
discussions, and the organizers of the Summer Institute 2000, Japan,
in which a part of this work was carried out.

\newcommand{\J}[4]{{\sl #1} {\bf #2} (#3) #4}
\newcommand{\andJ}[3]{{\bf #1} (#2) #3}
\newcommand{\AP}{Ann.\ Phys.\ (N.Y.)}
\newcommand{\MPL}{Mod.\ Phys.\ Lett.}
\newcommand{\NP}{Nucl.\ Phys.}
\newcommand{\PL}{Phys.\ Lett.}
\newcommand{\PR}{Phys.\ Rev.}
\newcommand{\PRL}{Phys.\ Rev.\ Lett.}
\newcommand{\PTP}{Prog.\ Theor.\ Phys.}
\newcommand{\hepth}[1]{{\tt hep-th/#1}}


\begin{thebibliography}{99}

\bibitem{sw}
 N.\ Seiberg and E.\ Witten, ``{\it String Theory and 
 Noncommutative Geometry}'',
 \hepth{9908142}, \J{JHEP}{9909}{1999}{032}

\bibitem{gmss}
 R.\ Gopakumar, S.\ Minwalla, N.\ Seiberg and A.\ Strominger, 
 ``{\it OM Theory in Diverse Dimensions}'',
 \hepth{0006062}, \J{JHEP}{0008}{2000}{008}

\bibitem{bbss1}
 E.\ Bergshoeff, D.\ S.\ Berman, J.\ P.\ van der Schaar and
 P.\ Sundell ``{\it A Noncommutative M-Theory Five-brane}'',
 \hepth{0005026}

\bibitem{bbss2}
 E.\ Bergshoeff, D.\ S.\ Berman, J.\ P.\ van der Schaar and
 P.\ Sundell ``{\it Critical fields on the M5-brane and noncommutative 
 open strings}'',
 \hepth{0006112}

\bibitem{hlw}
 P.\ S.\ Howe, N.\ D.\ Lambert and P.\ C.\ West, ``{\it The Self-Dual
 String Soliton}'',
 \hepth{9709014}, \J{\NP}{B515}{1998}{203}

\bibitem{lm}
 D.\ L\"{u}st and A.\ Miemiec, ``{\it Supersymmetric M5-branes with
 $H$-field}'',
 \hepth{9912065}, \J{\PL}{B476}{2000}{395}

\bibitem{hsw}
 P.\ S.\ Howe, E.\ Sezgin and P.\ C.\ West, ``{\it Covariant 
 Field Equations of the M Theory Five-Brane}'',
 \hepth{9702008}, \J{\PL}{B399}{1997}{49}

\bibitem{apps}
 M.\ Aganagic, J.\ Park, C.\ Popescu and J.\ H.\ Schwarz, 
 ``{\it Worldvolume action of the M-theory five-brane}'',
 \hepth{9701166}, \J{\NP}{B496}{1997}{191}

\bibitem{blnpst1}
 I.\ Bandos, K.\ Lechner, A.\ Nurmagambetov, P.\ Pasti, D.\ Sorokin 
 and M.\ Tonin, ``{\it Covariant Action for a D=11 Five-Brane with
 the Chiral Field}'',
 \hepth{9701037}, \J{\PL}{B398}{1997}{41}

\bibitem{blnpst2}
 I.\ Bandos, K.\ Lechner, A.\ Nurmagambetov, P.\ Pasti, D.\ Sorokin 
 and M.\ Tonin, ``{\it Covariant Action for the Super-Five-Brane
 of M-Theory}'',
 \hepth{9701149}, \J{\PRL}{78}{1997}{4332}

\bibitem{cns}
 M.\ Cederwall, B.\ E.\ W.\ Nilsson and P.\ Sundell, 
 ``{\it An action for the super-5-brane in $D=11$ supergravity}'',
 \hepth{9712059}, \J{JHEP}{9804}{1998}{007}

\bibitem{blnpst3}
 I.\ Bandos, K.\ Lechner, A.\ Nurmagambetov, P.\ Pasti, D.\ Sorokin 
 and M.\ Tonin, ``{\it On the Equivalence of Different Formulations 
 of the M Theory Five-Brane}'',
 \hepth{9703127}, \J{\PL}{B408}{1997}{135}

\bibitem{rtcghn}
 J.\ G.\ Russo and A.\ A.\ Tseytlin
 ``{\it Waves, boosted branes and BPS states in M-Theory}'',
 \hepth{96011047}, \J{\NP}{B490}{1997}{121} ;
 M.\ Cederwall, U.\ Gran, M.\ Holm and B.\ E.\ W.\ Nilsson
 ``{\it Finite Tensor Deformations of Supergravity Solitons}'',
 \hepth{9812144}, \J{JHEP}{9902}{1999}{003}

\bibitem{bs}
 D.\ S.\ Berman  and P.\ Sundell
 ``{\it Flowing to a noncommutative (OM) five brane via its
 supergravity dual}'',
 \hepth{0007052}

\bibitem{m}
 K.\ Hashimoto and T.\ Hirayama, ``{\it Branes and BPS Configurations
 of Non-Commutative/Commutative Gauge Theories}'',
 \hepth{0002090} ;
 S.\ Moriyama, ``{\it Noncommutative Monopole from Nonlinear 
 Monopole}'',
 \hepth{0003231}, \J{\PL}{B485}{2000}{278}

\bibitem{h}
 K.\ Hashimoto, ``{\it Born-Infeld Dynamics in Uniform Electric 
 Field}'',
 \hepth{9905162}, \J{JHEP}{9911}{1999}{005} ;
 D.\ Mateos, ``{\it Non-commutative vs. Commutative Descriptions of
 D-brane BIons}'',
 \hepth{0002020}, \J{NP}{B577}{2000}{139}

\bibitem{cm}
 C.\ Callan and J.\ Maldacena
 ``{\it Brane Dynamics From the Born-Infeld Action }'',
 \hepth{9708147}, \J{\NP}{B513}{1998}{198}

\end{thebibliography}
\end{document}